\documentclass{aa}  

\usepackage{graphicx}
\usepackage{natbib}
\usepackage{amsmath}
\usepackage{xcolor}

\newcommand{\msun}{$M_{\odot}$}
\newcommand{\rsun}{$R_{\odot}$}
\newcommand{\mearth}{$M_{\oplus}$}
\newcommand{\rearth}{$R_{\oplus}$}

\newcommand{\kms}{${\rm km\;s^{-1}}$}
\newcommand{\ms}{${\rm m\;s^{-1}}$}
\newcommand{\cms}{${\rm cm\;s^{-1}}$}
\newcommand{\ma}{${\rm m{\AA}}$}

\usepackage{txfonts}
\begin{document} 

   \title{The solar gravitational redshift from HARPS-LFC \\ Moon spectra
   \thanks{Based on observations taken with the ESO 3.6 m telescope at La Silla 
   Observatory, Chile.}
   \fnmsep 
   \thanks{Tables A.1 and A.2 are only available in electronic form at the CDS via 
   anonymous ftp to cdsarc.u-strasbg.fr (130.79.128.5)
   or via http://cdsweb.u-strasbg.fr/cgi-bin/qcat?J/A+A/}
   }

   \subtitle{A test of the general theory of relativity}

   \author{J. I. Gonz\'alez Hern\'andez\inst{1,2}
         \and
         R. Rebolo\inst{1,2,3}
         \and
         L. Pasquini\inst{4}
         \and
         G. Lo Curto\inst{4}
         \and
         P. Molaro\inst{5,6}
         \and
         E. Caffau\inst{7}
         \and
         H.-G. Ludwig\inst{8,7}
         \and
         M. Steffen\inst{9}
         \and
         M. Esposito\inst{10}
         \and
         A. Su\'arez Mascare\~no\inst{1,2}
         \and
         B. Toledo-Padr\'on\inst{1,2}
         \and 
         R. A. Probst\inst{11,12}
         \and
         T. W. H{\"a}nsch\inst{11}
         \and
         R. Holzwarth\inst{11,12}
         \and
         A. Manescau\inst{4} 
         \and
         T. Steinmetz\inst{12}
         \and
         Th. Udem\inst{11}
         \and 
         T. Wilken\inst{10,11}
        }

\institute{
Instituto de Astrof\'{\i}sica de Canarias, 
V\'{\i}a L\'actea, 38205 La Laguna, Tenerife, Spain\\
\email{jonay@iac.es}
         \and
         Universidad de La Laguna, Departamento de Astrof\'{\i}sica, 
         38206 La Laguna, Tenerife, Spain
         \and
         Consejo Superior de Investigaciones Cient\'{\i}ficas, E-28006 Madrid, Spain
         \and
         European Southern Observatory, Karl Schwarzschild Strasse 2, 85748 
         Garching, Germany
         \and
         INAF Osservatorio Astronomico di Trieste, via G. B. Tiepolo 11, 34143 
         Trieste, Italy
         \and
         Institute for Fundamental Physics of the Universe  (IFPU) , Via Beirut 2, 
         34151, Grignano, Trieste, Italy
         \and
         GEPI, Observatoire de Paris, Universit{\'e} PSL, CNRS, 5 Place Jules 
         Janssen, 92190 Meudon, France
         \and
         Zentrum f{\"u}r Astronomie der Universit{\"a}t Heidelberg, 
         Landessternwarte, K{\"o}nigstuhl 12, 69117 Heidelberg, Germany
         \and
         Leibniz-Institut f{\"u}r Astrophysik Potsdam (AIP), An der Sternwarte 16, 
         14482 Potsdam, Germany
         \and
         Th{\"u}ringer Landessternwarte Tautenburg, Sternwarte 5, 07778 
         Tautenburg, Germany
         \and
         Max-Planck-Institut f{\"u}r Quantenoptik, Hans-Kopfermann-Str. 1, 85748 
         Garching, Germany
         \and
         Menlo Systems GmbH, Am Klopferspitz 19a, 82152 Martinsried, Germany
}

   \date{Received June 10, 2020; accepted XX 15, 2020}

% \abstract{}{}{}{}{} 
% 5 {} token are mandatory
 
  \abstract
   {The general theory of relativity predicts the redshift of spectral lines 
   in the solar photosphere as a consequence of the gravitational potential of the Sun. 
   This effect can be measured from a solar disk-integrated flux spectrum of the Sun's 
   reflected light on Solar System bodies.}
   {The laser frequency comb (LFC) calibration system attached to the HARPS 
   spectrograph offers the possibility of performing an accurate measurement of the solar 
   gravitational redshift (GRS) by observing the Moon or other Solar System bodies. 
   Here, we analyse the line shift observed in Fe absorption lines from five 
   high-quality HARPS-LFC spectra of the Moon.}
   {We selected an initial sample of 326 photospheric Fe lines in the spectral range 
   between 476--585~nm and measured their line positions and equivalent widths (EWs). 
   Accurate line shifts were derived from the wavelength position 
   of the core of the lines compared with the laboratory wavelengths of Fe lines. 
   We also used a CO$^5$BOLD 3D hydrodynamical model atmosphere of the Sun to compute 
   3D synthetic line profiles of a subsample of about 200 spectral Fe lines centred 
   at their laboratory wavelengths. 
   We fit the observed relatively weak spectral Fe lines (with EW$<180$~\ma) with 
   the 3D synthetic profiles.}
   {
   Convective motions in the solar photosphere do not affect the line cores of
   Fe lines stronger than about $\sim 150$~\ma.
   In our sample, only 15 \ion{Fe}{i} lines have EWs in the range 
   $150 <$~EW(\ma)~$< 550$, providing a measurement of the solar GRS 
   at $639\pm14$~\ms, which is consistent with the expected theoretical value on 
   Earth of $\sim 633.1$~\ms.
   A final sample of about 97 weak Fe lines with EW~$<180$~\ma~allows us to 
   derive a mean global line shift of $638\pm6$~\ms , which is in agreement with the 
   theoretical solar GRS.
   }
   {These are the most accurate measurements of the solar GRS obtained thus far.   
   Ultrastable spectrographs calibrated with the LFC over a larger spectral range, 
   such as HARPS or ESPRESSO, together with a further improvement on the 
   laboratory wavelengths, could provide a more robust measurement of the 
   solar GRS and further testing of 3D hydrodynamical models.}


   \maketitle
%
%-------------------------------------------------------------------

\section{Introduction} \label{sec:intro}

The general theory of relativity~\citep{ein11,ein16} predicts the gravitational 
redshift of spectral lines in the solar photosphere. However, the  
effects of gravity on light were first demonstrated by light deflection during 
a solar eclipse and by its magnified effects in white dwarfs~\citep{gre72}. 
In the Sun, the gravitational redshift is competing with the blueshift of the 
lines due to the convective motions.
Convection produces granulation phenomena, which induce line shifts which are 
opposite to the gravitational redshift. The granulation pattern is 
formed by upward flows with a velocity of 1-2~\kms in the inner parts of the granules 
and downward flows in the intergranular lanes, which are faster 
by a factor of between two and three~\citep{dra81,dra90}.
The granulation pattern has a scale size of 0.6-1.3 Mm and a lifetime 
of about 8-12 min~\citep{spr90,hir99,sch00}. 
Larger scale sizes of 15-30 and 5-10 Mm and lifetimes of 20-30 hr and $\sim 2$~hr
are also present, and referred to as supergranulation and mesogranulation, which produce 
vertical raising velocities of 40-60~\ms, respectively~\citep{sch00},
with larger horizontal velocities and more relevant, perhaps, towards the solar limb.

This complex granulation pattern produces asymmetric disk-integrated photospheric lines 
with observed Doppler blue shifts down to $\sim -500$~\ms~\citep{all98a,rei16} after 
subtracting the  theoretical value of the solar gravitational redshift. 
Granulation is also variable and on timescales larger than 1~hr could produce 
radial velocity (RV) variations of $\sim 1$~\ms~\citep{meu15}. In addition, the 
$\sim 5$~min solar oscillations are associated with a vertical 
velocity component of $\sim 0.4$~\kms~\citep{kei78}, producing a RV signature in
disk-integrated light of only 0.1-4~\ms~\citep{dum11}. 

Active regions, mainly spots and faculae, with a lifetime of 10-50 days, located at 
different positions of the  solar disk as the Sun rotates, can also yield RV 
signatures of about 0.4~\ms~due to the flux imbalance  and about 2.4~\ms~from the 
suppression of convective motions in magnetic regions~\citep{lag10,meu10a,hay16}. 
The inhibition of convection appears to be the dominant source of RV variations during 
the solar cycle, with an amplitude of 8-10~\ms~\citep{meu10a,meu10b,lan16}.

%Doppler shifts of photospheric lines as a signature of granulation can be measured 
%using solar line bisectors~\citep{dra81} and line core positions~\citep{all98a}. 
%Bisectors also provide a description of the asymmetry of the lines~\citep{gra09}. 
The modelling of granulation has been done using empirical models that are relatively 
simple, consisting of between two and four components 
\citep{dra81,dra90} or complex hydrodynamical simulations, including
magnetic fields~\citep[e.g.][]{ceg13}. Hydrodynamical simulations have also been able 
to fairly reproduce the shapes and absolute shifts of \ion{Fe}{i} lines~\citep{asp00}. 
The convective blueshift caused by the granulation pattern can be explained as 
flux-dominant rising granules whose velocity diminishes with height. 
Strong and weak lines form generally in the outer and inner layers, respectively, 
of the photosphere~\citep{mag86,gro94,gra05}.  
Thus, cores of stronger lines forming in outer layers of the photosphere show 
smaller convective blueshifts than weaker lines forming deeper in the photosphere. 
Therefore, the Doppler shifts of strong lines may, in principle, mostly be 
associated to the gravitational redshift.

The accuracy of line positions in the solar spectra has been extensively investigated 
and some solar line atlases have been published in recent 
decades~\citep[e.g.][]{all98b,mol12}, using Fourier transform spectrometer (FTS) 
observations of the Sun~\citep{kur84,wal11}. 
The wavelength positions of solar lines in the Kitt Peak FTS spectra~\citep{kur84} 
have an uncertainty of 50-150~\ms~and zero offset of about 100~\ms~\citep{all98b,mol11}. 
The line precision and accuracy has been improved to 45-90~\ms~with HARPS observations 
of solar spectra obtained from asteroids~\citep{mol11,mol12,lan15}. 
Using HARPS spectra of the Moon wavelength calibrated with the laser frequency comb, 
\citet{mol13} produced a  solar atlas in the region 480-580 nm with a 
mean accuracy of $\sim 12$~\ms. \citet{rei16} obtained a new solar atlas 
from observations of disk-integrated FTS spectra. The radial velocity is consistent 
with that of the HARPS-LFC  at the level of few \ms, thus confirming the systematic 
offset of 100~\ms~of the Keat Peak solar spectrum of~\citet{kur84,wal11}.

The general theory of relativity~\citep{ein11,ein16} predicts the gravitational 
redshift of spectral lines in the solar photosphere as
$v_{\rm GRS,\odot}= (G M_\odot / c R_\odot)$, where $G$ is the gravitational constant, 
$c$ is the speed of light, and \msun\ and \rsun\ are the mass and radius of the Sun, 
respectively. 
Using the nominal values of the astronomical constants recommended by the 
IAU~\citep{prs16}, this equation gives $v_{\rm GRS,\odot}=636.31$~\ms. 
The same effect from the solar photosphere to 1 AU~\citep{all09,roc14} amounts to:
\begin{equation*}
v_{\rm GRS,1{\rm AU}}=\frac{G M_\odot}{c}\left(\frac{1}{R_\odot}-
\frac{1}{\rm 1AU}\right)=633.35{\rm \;m\;s}^{-1}.
\end{equation*}
The distance of the Earth on 25 November 2010 was of 147.75 millions km, 
which leads to a small correction of about +4~\cms. 
The gravitational effect of the Moon is zero since it is a reflection, however, 
in accounting for the gravitational potential of the Earth, 
$v_{\rm GRS,\oplus}= (G M_\oplus / c R_\oplus)=0.21$~\ms, where \mearth~and \rearth~ 
are the nominal mass and radius of the Earth, we get a final value of:
\begin{equation*}
v_{\rm GRS,{\rm theo}}=v_{\rm GRS,1{\rm AU}}-v_{\rm GRS,\oplus}=633.10{\rm \;m\;s}^{-1}.
\end{equation*}
The solar GRS effect is quite large and it may seem rather surprising that it has never 
been measured with high accuracy.  
We can roughly divide the attempts in two categories: observations 
of single lines in resolved regions of the solar disk and observations of the 
integrated solar disk, usually done with stellar spectrographs. 
The first class of observations usually demonstrate quite a high level of precision, 
but they are limited by the fact that, with the exception of sunspot umbrae, each 
fraction of the solar surface is subject to peculiar velocities~\citep{bec77}.

Using the solar disk-integrated atlases, various authors have provided an observational 
measurement of the solar GRS from strong lines in the solar photosphere, with values 
close to the theoretical prediction: $v_{\rm GRS,{\rm obs}}=612\pm58$~\ms~\citep{all98a}, 
$v_{\rm GRS,{\rm obs}}=601\pm43$~\ms~\citep{mol12}. Using solar spectra in the small 
spectral region 5188–5212~{\AA} calibrated with an iodine-cell, \citet{tak12} obtained
a result of $v_{\rm GRS,{\rm obs}}=698\pm133$~\ms. More recently, \citet{roc14} presented 
disk-integrated sunlight observations centred on the \ion{K}{I}~7699~{\AA} line with a 
spectral width of 15~{\AA} over an extended period, between 1976–2013, to derive 
$v_{\rm GRS,{\rm obs}}=600.4\pm0.8$~\ms,
with an amplitude of $\pm 5$~\ms, which is which is in anti-correlation with regard to 
the solar magnetic cycle.

Resolved observations suffer from the peculiar motions on the solar surface, whereas 
integrated disk observations suffer from the limited accuracy of traditional 
spectrographs. 
With the advent of a new generation of exoplanet hunters, such as HARPS, and the use of 
calibration systems based on laser frequency combs, solar observations are now capable 
of reaching the required level of accuracy.

The laser frequency comb (LFC) is a new wavelength calibration system for high-resolution 
astronomical spectrographs which has been tested on the HARPS 
spectrograph~\citep{wil10,wil12}. In terms of the determination of individual line 
positions, Th-Ar lamps can achieve an accuracy on the level  of several tens of 
\ms~\citep{loc12,pal83}. 
When accounting for the several thousand lines available in the spectral range of 
HARPS, the overall precision can be increased by about a factor of 100, with 
limitations related to the blending of lines, non-uniform density of Th lines across the
spectrum, and lamp aging. 
We should note that for determining the solar GRS, we need to measure the absolute
wavelength of spectral lines and, therefore, wavelength accuracy (rather than precision)
is the most relevant parameter. 
The comparison of line positions of LFC and Th-Ar calibrated spectra in the HARPS 
spectrograph have revealed S-type distortions on each order 
along the whole spectral range with an amplitude of $\pm 40$~\ms~and whose precise 
shape is also wavelength-dependent~\citep{mol13}. Several test campaigns of the LFC
coupled to HARPS have demonstrated that its calibration can give a short-term 
repeatability of 2.5~\cms~\citep{wil12}, while Th-Ar lamps can only reach a 
photon noise level of 10~\cms. In addition, the RV precision of Th-Ar lamps is about 
30~\cms~\citep[e.g.][]{loc10}.
Very recently, \citet{pro20} demonstrated repeatability of the 
line positions of the upgraded HARPS LFC system down to the photon noise level 
of 1~\cms, as compared with the 10~\cms~level reached by Th-Ar lamps, 
opening up a new horizon for current and future ultra-stable high-resolution 
spectrographs, along with bringing new science frontiers closer, such as the 
detection of Earth twins orbiting Sun-like stars in their habitable zone. 
The same authors have also shown that wavelength calibrations obtained with two 
separate LFCs on HARPS were consistent to better than 60~\cms~\citep[see also][]{mil20}.
In this work, we exploit the high accuracy of the LFC wavelength solution using LFC 
calibrated HARPS spectra of the Sun's light reflected on the Moon~\citep{mol13} to 
measure the solar gravitational redshift as a test of the general theory of 
relativity~\citep{ein16}. 

\section{Observations and data analysis} \label{sec:obs}

We performed observations of the Moon on 25 November 2010 with the HARPS
spectrograph %wavelength calibrated with the LFC, 
installed at the 3.6m-ESO telescope in the {\it Observatorio de La Silla} 
(see Table~\ref{tab:obs}). 
HARPS is a cross-dispersed echelle, high-resolution spectrograph fed with two 
fibres of 1-arcsec aperture on sky (fiber A: science and fiber B: calibration), 
which follow approximately the same optical path~\citep{may03}. 
The HARPS fibers are equipped with a double scrambler to provide 
both image and pupil stability, which is an important difference with respect to
spectrographs fed through a slit, where a non-uniform slit illumination induces
radial velocity differences between different exposures.
HARPS is an ultra-stable instrument with the capacity to achieve both short-term and 
long-term precision below 0.8~\ms~\citep{pep11,pep14}. 
The calibration on fiber B is used to track the instrument drift, typically below 
$\pm 1$~\ms, during night observations with respect to afternoon 
calibrations~\citep{may03}.

\begin{table}[!ht]
\centering
\caption{Log of observations\label{tab:obs}}
%\begin{tabular}{cD@{$\pm$}D}
\begin{tabular}{lrrrr}
%\tablewidth{0pt}
\hline
\hline
%Parameter & \multicolumn2c{Value} & \multicolumn2c{Uncertainty}\\
UT(start) & Label & $T_{\rm exp}$ & RV$_c$ & S/N \\
 h:m:s &  & s & \kms & \\
\hline
05:47:20.802 & s802 & 60  &  $-1.190$ & 241-286-330 \\
05:49:42.429 & s429 & 150 &  $-1.187$ & 383-457-524 \\
05:53:19.301 & s301 & 60  &  $-1.182$ & 245-293-337 \\
05:55:20.898 & s898 & 120 &  $-1.179$ & 371-443-509 \\
05:57:54.476 & s476 & 120 &  $-1.176$ & 383-457-526 \\
\hline
\multicolumn{5}{l}{NOTE. - UT at start, exposure times, $T_{\rm exp}$, expected radial} \\
\multicolumn{5}{l}{velocity computed with JPL ephemerides, RV$_c$, signal to} \\
\multicolumn{5}{l}{noise at centre of order 34, 45 and 57 of the HARPS-LFC} \\
\multicolumn{5}{l}{observations of the Moon carried out on 25 November 2010.} \\
%\hline
\end{tabular}
\end{table}

HARPS provides a median resolving power of $R=\lambda/\delta\lambda\sim115,000$ 
covering the spectral range of 380--690~nm distributed over 72 spectral echelle 
orders, with a small gap at 530-539~nm. The LFC consists of thousands of lines that are equally spaced in frequency and can be 
described with a simple equation $f= f_{\rm off} + n f_{\rm rep}$, where 
$f_{\rm rep}$ is the repetition frequency (that is, the frequency difference between 
adjacent lines) and $f_{\rm off}$ is the offset frequency, and $n$ is an integer 
that is commonly referred to as the mode number. 
The LFC is linked to an atomic clock which 
ensures an extremely high accuracy (better than 1 m/s) and long-term 
stability~\citep{wil12,loc12}.
The LFC, as described in \citet{wil12}, uses three concatenated Fabry-Perot cavities 
to filter the unwanted modes to a line spacing of $f_{\rm rep}=18$~GHz. This is followed 
by a frequency doubling of the LFC spectrum to centre it at about 525 nm
and subsequent spectral broadening in a tapered photonic crystal fibre. 
Thus, the LFC is imaged onto both HARPS blue and red chips covering a wavelength 
range of about 125 nm, spanning from 465 to 590 nm, from spectral order \#34 to \#57 
of the 72 orders of HARPS.

The LFC offers about 350 lines (usually called modes) per spectral order equally 
spaced in frequency and with similar intensity (as a comparison,  Th-Ar lamps offer 
less than 100 lines per spectral order with very different intensities, sometimes 
blended or even saturated). 
This allows for wavelength calibrations of a fraction of an order and we separately calibrated 
 each of the eight master lithographic blocks of 512 pixels in the dispersion 
direction~\citep{mol13}. HARPS blue and red chips are, in fact, built up with 16 master 
blocks of 1024 x 512 pixels. The use of the LFC allowed us to
discover the S-type distortions and up to about 70~\ms~`jumps' in the wavelength 
solutions between adjacent master blocks associated to different pixel sizes along 
the stitching borders of these blocks~\citep{wil10,mil20}. 

\begin{figure}
\begin{center}
{\includegraphics[clip=true,width=95mm,angle=0]{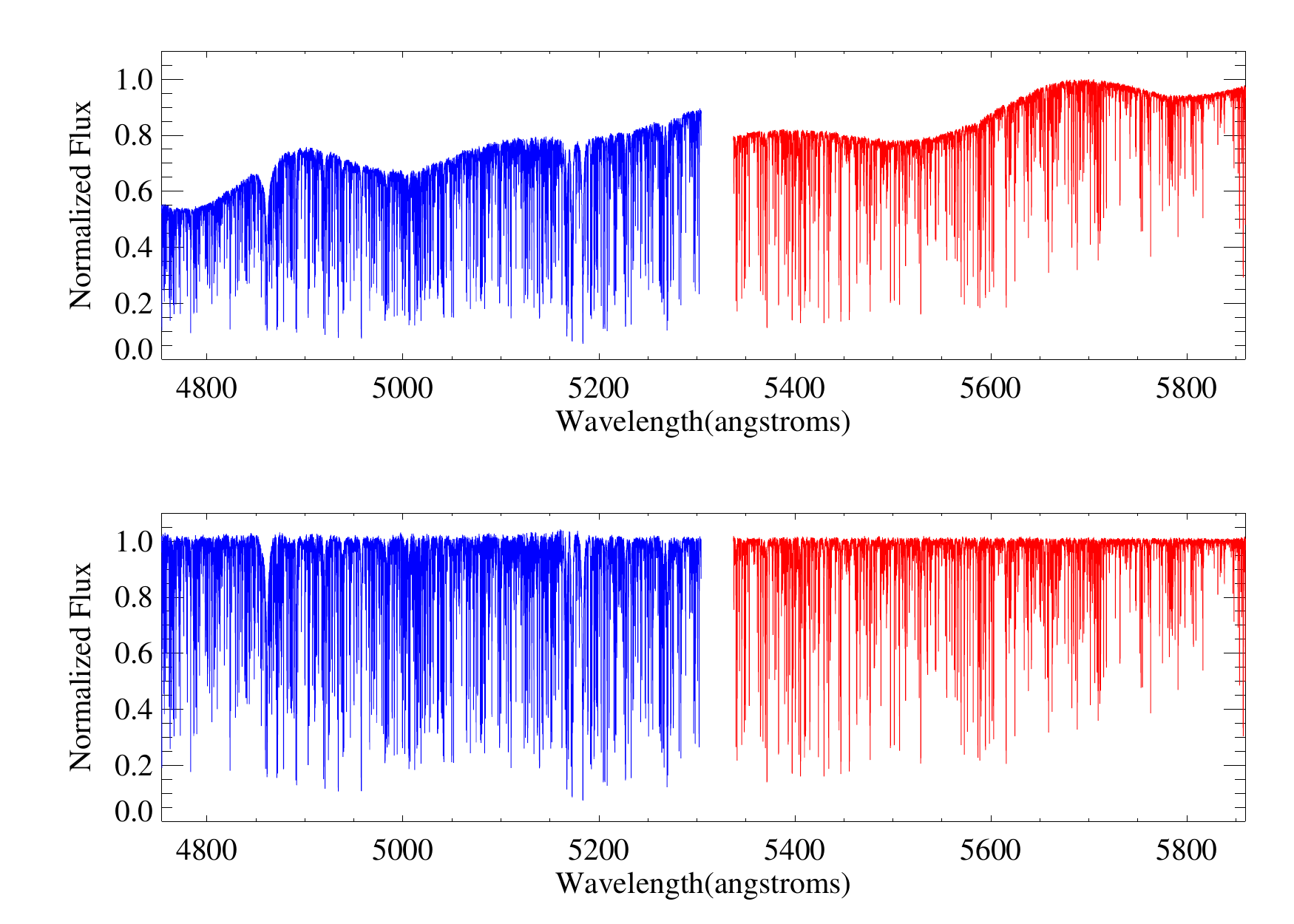}}
\end{center}
\caption{
{\it Top panel}: Blaze-corrected, co-added, merged HARPS-LFC spectrum of the Moon 
for both blue and red chips of HARPS. 
{\it Bottom panel}: Normalised (order-by-order), merged HARPS-LFC spectrum.
\label{fig:spec}
}
\end{figure}

HARPS-LFC spectra of the Sun's reflected light on the Moon were collected pointing 
to the Moon's centre. During these observations, 86\% of the Moon was illuminated.
Five HARPS spectra of the Moon on fiber A were taken with simultaneous LFC calibration 
on fiber B, providing a simultaneous reference spectrum to monitor 
overnight instrumental drifts. The instrumental drift is computed from the displacement 
in pixels of the LFC spectrum acquired on fiber B (simultaneously with the Moon 
spectra on fiber A) with respect to a reference LFC spectrum. For each LFC mode the 
displacement is weighted by the signal collected on that mode.
The HARPS-LFC observations cover a time span of 10 min, ensuring that solar oscillations 
should be averaged out. The exposure times varies between 60 to 150 seconds, providing 
signal-to-noise ratios (S/N) in the range between $\sim 290-460$ (see Table~\ref{tab:obs}).  
Each spectrum was reduced using the standard HARPS pipeline and the LFC frames were
calibrated using eight third-degree polynomials corresponding to the eight master blocks 
of 512 pixels. 
These LFC wavelength calibrated 2D order spectra were blaze-corrected using 
the blaze function of each order. We then normalised these 2D spectra order by order 
with a third-order polynomial using our own automated IDL-based routine. The normalised
spectra after order merging were rebinned to a constant wavelength step of 
0.01~{\AA}~pixel$^{-1}$ equivalent to $\sim 0.8$~\kms~pixel$^{-1}$. 
Each spectrum was corrected for radial velocity (see Table~\ref{tab:obs}) 
using the JPL ephemeris to place them in the laboratory rest frame~\citep[see][for 
further details]{mol13}. The computed radial velocity 
accounts for the motions of the Moon with respect to the observer and 
the Sun, and also includes a component of 3.2~\ms~due to the rotation of the Moon, 
since the sunlight impacts the receding lunar hemisphere before being reflected 
at an angle of $43^{\circ}$ at the time of the observation.
In Fig.~\ref{fig:spec} we display the merged, normalised spectrum one of the five Moon 
HARPS-LFC spectra. A proper normalisation procedure is very relevant for the 
determination of the line core centres of individual Fe lines. 

\begin{figure}
\begin{center}
{\includegraphics[clip=true,width=95mm,angle=0]{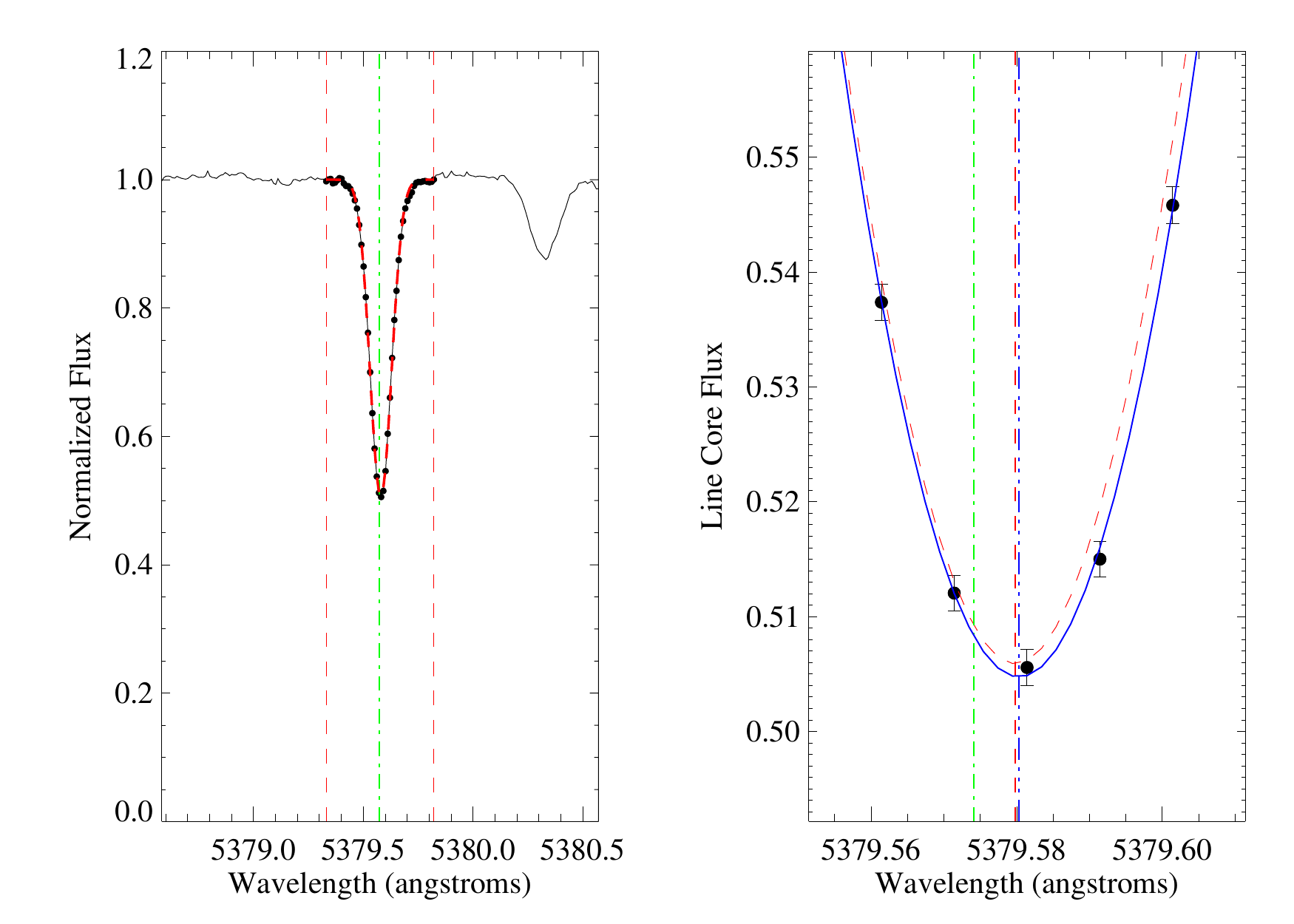}}
\end{center}
\caption{ 
Example of the line core fit of the line \ion{Fe}{i} $\lambda_{\rm lab}$ 
5379.5740~{\AA} described in Section~\ref{sec:lc}.
{\it Left panel}: Normalised flux of a HARPS-LFC spectrum (black thin line) of 
the Moon, labelled s429 in Table~\ref{tab:obs}, centred on $\lambda_{\rm lab}= 
5379.5740$~{\AA} together with a single Gaussian fit (red dashed line) 
in a region of 0.5~{\AA} (vertical red dashed lines) centred on the 
laboratory wavelength (vertical green dashed-dotted line). 
{\it Right panel}: Parabolic fit (blue solid line) to five points of the line 
core compared to the single Gaussian fit (red dashed line). 
Vertical lines show the wavelength position of the centroids of the 
parabolic fit (blue dashed-three dotted line) and 
the Gaussian fit (red dashed line), together with the laboratory wavelength 
(vertical green dashed-dotted line).
\label{fig:lcfit}
}
\end{figure}

\section{Line core shifts \label{sec:lc}} 

We selected Fe lines from the linelists in previous works~\citep{sou07,mol13} available 
in the spectral range 465--590 nm, complemented with Fe lines extracted from the Vienna 
Atomic Line Database~\citep[VALD\footnote{Vienna Atomic Spectra Database
available at \\
{http://vald.astro.uu.se}}][]{pis95}, using the extract stellar 
option with solar parameters.
We search for apparently isolated \ion{Fe}{i} and \ion{Fe}{ii} lines by visually 
inspecting the HARPS-LFC spectrum of the Moon with the IRAF package {\sc splot}. 
The atomic line parameters of Fe lines were adopted from VALD, including wavelength, 
excitation potential, and oscillator strength. We also collect, for these Fe lines, the 
laboratory wavelengths from \citet{nav94,nav12,nav13}. 
We note that the laboratory wavelengths of \ion{Fe}{i} lines with a quality flag 
labelled with A have an accuracy in the range $0.28-1.25$~\ma, corresponding 
to $16-75$~\ms~\citep{nav94}.
Those lines flagged as B, C, D have uncertainties in the range
$0.4-2.5$~\ma, $0.8-5.0$~\ma, and $>5.0$~\ma, corresponding to $24-150$~\ms,
$48-300$~\ms,$>300$~\ms~\citep{nav94}. 
The final linelist contains 326 Fe lines, 200 in the blue CCD and 126 in the 
red CCD. Among these lines there are 152, 42, 3, 3 lines in the blue chip 
and 101, 15, 3, 7 lines in the red chip with flags A, B, C, D, respectively.
There are only four \ion{Fe}{ii} lines in the blue chip that have uncertainties in 
the range of $0.2-0.9$~\ma, i.e. $12-51$~\ms~\citep{nav13}.
 
\citet{nav11} discussed the calibration of wavenumber measurements used to
infer the original published wavelengths of Fe lines~\citep{nav94}. 
They used 28 \ion{Ar}{ii} lines as wavenumber standards based on the results in 
\citet{nor73}, that were subsequently remeasured by \citet{wha95}, providing, in 
principle, a higher precision and accuracy. They suggested to increase the wavenumber
measurements of \ion{Fe}{i} by $6.7\pm0.8$ parts in $10^8$. 
We have downloaded, from the NIST Atomic Spectra 
Database\footnote{NIST Atomic Spectra Database
(ver. 5.7.1) available at \\
{https://physics.nist.gov/asd}}, the  wavelengths 
of Fe lines, together with the Ritz wavelengths derived from the recalibrated 
energy levels, both with the same constant quantity of $6.7$ parts 
in $10^8$~\citep{NIST_ASD}. 
These wavelengths are shorter, on average, by about 20~and 35~\ms, respectively, 
compared to the original wavelengths in \citet{nav94}. 
The mean uncertainties on wavelengths also extracted from 
NIST database are 1.1~and 0.4~\ma\ corresponding to 64 and 23~\ms.
Throughout this work, we use as our laboratory wavelengths, $\lambda_{\rm lab}$, 
the original wavelengths in \citet{nav94} and we discuss in Section~\ref{sec:grs} how 
the results change when using the recalibrated wavelengths, 
$\lambda_{\rm nist}$, and the recalibrated Ritz wavelengths, $\lambda_{\rm ritz}$, 
%as given in Table~\ref{tab:lc}.
as given in Table~A.1.

The equivalent widths (EWs) of the 326 lines were measured with the 
{\sc TAME}~\citep{kan12} 
code\footnote{TAME can be downloaded at \\
{http://astro.snu.ac.kr/$\sim$wskang/tame/} or at \\
{http://psychiee.byus.net/tame/} }, which
{\sc } uses a linelist extracted from VALD to identify nearby spectral lines 
that can affect the EW determination of a given spectral line and performs a 
multi-Gaussian fit to a spectral range of a few angstroms. 
For a few strong lines with EWs greater than 170~\ma~blended with weaker lines, we 
use the {\sc splot} tool within IRAF package to fit them using a Lorentzian 
function to account for the strong wings of those spectral lines. There were also
several weaker lines with some peculiarities for which {\sc TAME} fails to provide
a good fit, so we additionally measured their EWs with the {\sc splot} tool. A total of
six and nine Fe lines in the blue and red chip were measured with IRAF.

\begin{figure}
\begin{center}
{\includegraphics[clip=true,width=95mm,angle=0]{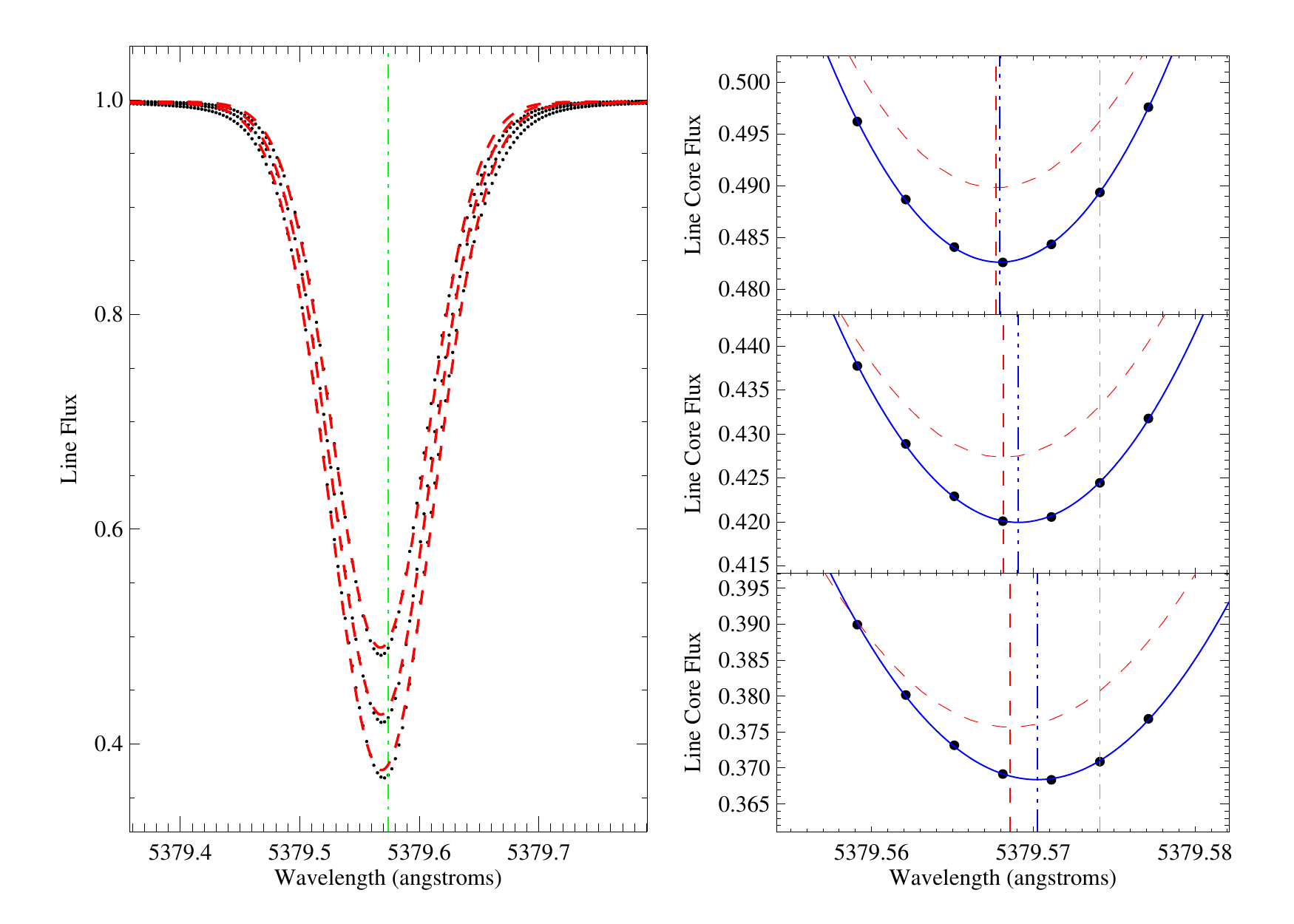}}
\end{center}
\caption{ 
Example of line core fit of the 3D profiles of the line 
\ion{Fe}{i} $\lambda_{\rm lab}$ 5379.5740~{\AA} described in Section~\ref{sec:lc3d}.
{\it Left panel}: Line 3D profiles for three
line strengths at solar abundance $\pm 0.1$~dex (black dots) together with single 
Gaussian fits (red dashed lines) and the laboratory wavelength 
(vertical green dashed-dotted line). 
{\it Right panels}: Parabolic fits
(blue solid line) to the line core of the three 3D profiles, together with single
Gaussian fits (red dashed lines). Vertical lines show the wavelength position of 
the centroids of the parabolic fit (blue dashed-three dotted line) and 
the Gaussian fit (red dashed line), together with the laboratory wavelength 
(vertical green dashed-dotted line).
\label{fig:lc3D}
}
\end{figure}

The {\sc TAME} output provides both a linelist with laboratory wavelengths and 
expected depths in the solar photosphere extracted from VALD and a linelist with 
the identified lines and fitted wavelengths and EWs. 
This allows us to identify possible line blends of \ion{Fe}{i},
\ion{Fe}{ii,} or other element species with wavelengths closer than 0.15~{\AA} and 
depths greater and 0.5 times the depth of the fitted Fe line.
With this rule, we classified 38 and 16 lines as blended Fe lines in the blue and red 
chips.
We visually inspected the output of the {\sc TAME} code to see if the multi-Gaussian 
fits were reasonable. In a few cases, we detected some Fe lines with blends that are not 
present in the linelists extracted from VALD and we classified these lines as problematic.
{\sc TAME} also performs an automatic local continuum fit to the highest flux point 
and, in some cases, due to surrounding weak lines or close-by very strong lines, this
procedure fails. We identified these lines and labelled them as problematic.
These 11 and 4 Fe lines in the blue and red chips, respectively, were also discarded.
Thus, a total of 69 Fe lines were discarded from the initial list of 326. 
Only two \ion{Fe}{ii} lines in the blue chip fulfilled these criteria.

\begin{figure}
\begin{center}
{\includegraphics[clip=true,width=95mm,angle=0]{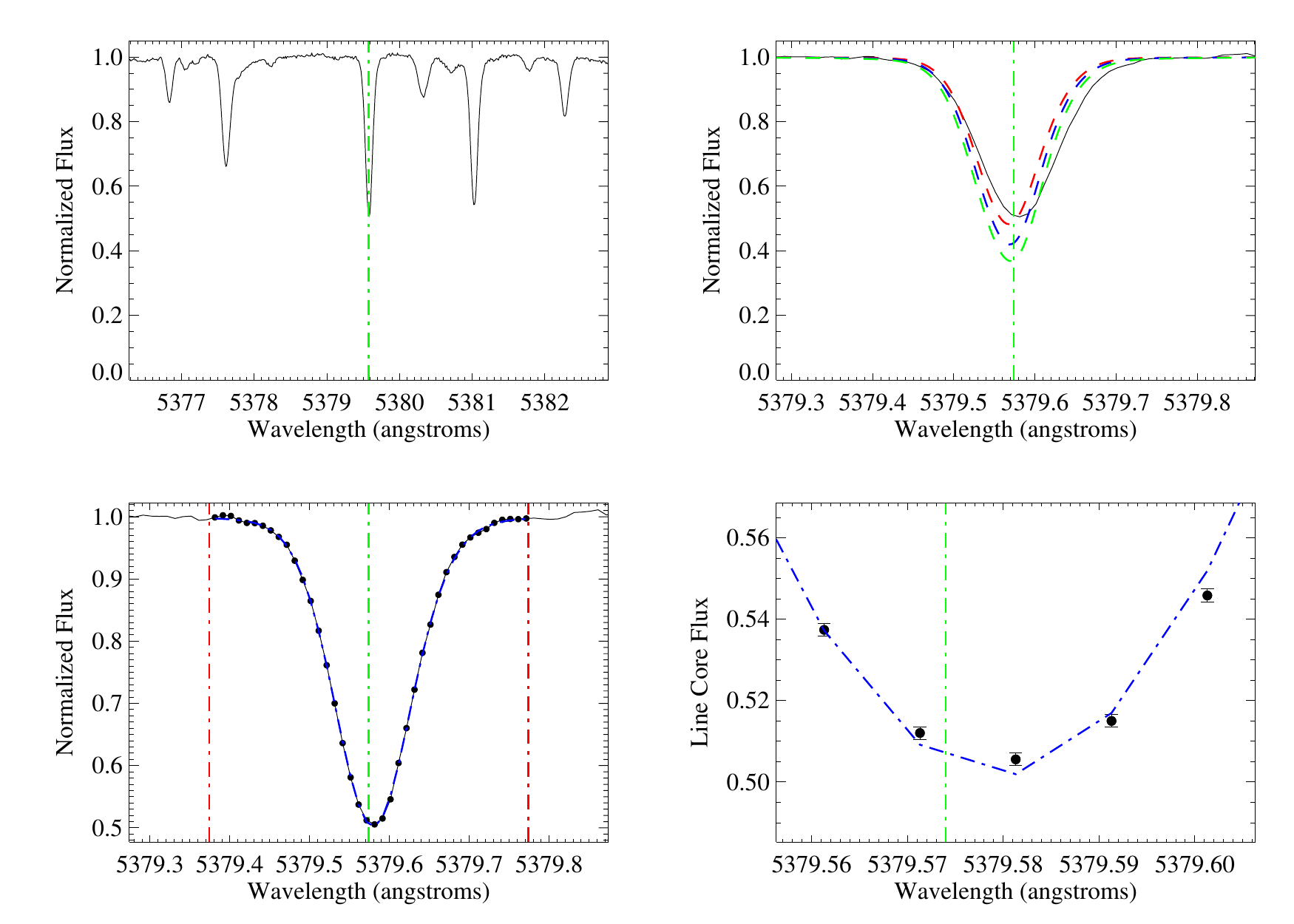}}
\end{center}
\caption{ 
Example of global line fit of the line \ion{Fe}{i} $\lambda_{\rm lab}$ 
5379.5740~{\AA} described in Section~\ref{sec:3dlfit}.
{\it Top-left panel}: Normalised flux of a HARPS-LFC spectrum (black thin line) 
of the Moon, labelled s429 in Table~\ref{tab:obs}, centred on 
$\lambda_{\rm lab}=5379.5740$~{\AA} as a vertical green dashed-dotted line. 
{\it Top-right panel}: Observed line profile together with
unbroadened 3D line profiles of three different strengths. 
{\it Bottom-left panel}: Global line fit (blue dashed-dotted line) to the line 
flux points within the spectral region of 0.4~{\AA} (region within the vertical 
red dashed-dotted lines) centred on the laboratory wavelength (vertical green 
dashed-dotted line). 
{\it Bottom-right panel}: Line core flux points (black dots) and global 
line fit (blue dashed-dotted line) with the laboratory wavelength 
(vertical green dashed-dotted line) shown as reference.
\label{fig:l3Dfit}
}
\end{figure}

We fitted the line cores of the remaining 257 Fe lines using our own automated IDL-based 
routine, shown in Fig.~\ref{fig:lcfit}. We identified the line cores  
with the laboratory wavelengths~\citep{nav94,nav12,nav13} of the spectral lines 
in a spectral range of $\sim$0.5~{\AA}. 
We fitted a single Gaussian to the line and selected the five flux points of the line core 
from the Gaussian centre and the minimum flux point of the line. 
We derived the centre of the line core from a parabolic fit to these
five flux points. In Fig.~\ref{fig:lcfit}, we display an example of the line core
fit to the observed \ion{Fe}{i} $\lambda_{\rm lab}$ 5379.5740~{\AA} line, with 
an EW of $60.2\pm1.2$~m{\AA}, using a parabolic fit that clearly provides a better 
centroid of the line core than a single Gaussian fit. 
The line core shift is computed as
$v_{\rm core,obs} = c (\lambda_{\rm core,obs}-\lambda_{\rm lab})/\lambda_{\rm lab}$, 
where $\lambda_{\rm lab}$ is the laboratory wavelength and $\lambda_{\rm core,obs}$ is 
the wavelength centroid of the parabolic fit. 
The centroid of the line core is clearly redshifted with
respect to the laboratory wavelength $\lambda_{\rm lab}=5379.5740$~{\AA} by 
$v_{\rm core,obs} =346.6\pm18.4$~\ms. 
We carried out the same exercise for each Fe line of each of the five HARPS-LFC 
spectra of the Moon. 
The value and uncertainty of the EW and of the line core shift are the mean and 
standard deviation, respectively, as resulting from the five measurements
of the five HARPS-LFC spectra.

\section{3D line profiles \label{sec:lc3d}} 

Synthetic 3D line profiles are based on a time-dependent, 3D, hydro-dynamical 
model atmosphere of the Sun computed with the CO$^5$BOLD~\citep{fre02} 
code\footnote{See details of CO$^5$BOLD code in \\
{http://www.lsw.uni-heidelberg.de/co5bold/workshop/cws\_intro.html} }.
The 3D model atmosphere has a box size of $5.6\times 5.6\times 2.27$ Mm$^3$, a 
resolution of $140\times 140\times 150$ ($N_X \times N_Y \times N_Z$) grid 
points~\citep{caf11}, and spans a 
range, in Rosseland optical depth, of about 
$-6.7 < \log \tau_{\rm Ross} < 5.5$ (from -1.4 Mm 
below to +0.9 Mm above $\tau_{\rm Ross} = 1$). 
We selected 20 representative snapshots from the full time sequence
to reduce the computing burden of the spectral synthesis calculations. These 
snapshots are equidistantly spaced in time, sufficiently separated in 
time to show little correlation, and cover 1.2 hours of solar time.
The 3D model atmosphere has a temporal average 
effective temperature of $\left<T_{\rm eff}\right>=5780$\,K, a surface gravity of  
$\log g=4.4$, and a metallicity of [Fe/H]$=0$ with 
A(Fe)\footnote{A(X)$=\log [N({\rm X})/N({\rm H})] + 12$ where $N({\rm X})$ 
represents the number density of nuclei of the element X.}=7.52.
The solar reference abundances are those given in \citet{caf11}.
The non-local radiative transfer is solved in 12 opacity bins on the standard 
grid of $140\times 140\times 150$ cells~\citep[see][for further details on 3D 
hydrodynamical model atmospheres]{lud09}.
The radiative heating terms are computed by solving the non-local LTE
transfer equation on a ray system of long characteristics for 12 opacity
bins.
The synthetic line profiles are computed with the line formation code 
{\sc Linfor3D}\footnote{http://www.aip.de/Members/msteffen/linfor3d}, 
taking into account the detailed 3D thermal structure and hydrodynamical
velocity field of the selected snapshots.
The adopted resolving power was 1666666, giving a velocity sampling of 
about 0.2~\kms. We use a vertical and three inclined angles and 
four azimuth angles~\citep{fre12}. 
We use the default broadening theory adopted in {\sc Linfor3D}, which
uses the standard Uns\"old theory that assumes a van der Waals interaction 
between the absorbing atom and the perturbing hydrogen atom, which, in principle,
could have an impact on the wings of strong lines but not on the core of the 
lines.
The statistical uncertainty on the wavelength position of the 
temporally and horizontally averaged 3D profiles is about 30~\ms,
estimated from the dispersion of individual snapshots and from the 
comparison with other solar models.

In Fig.~\ref{fig:lc3D}, we display an example of the 3D line profiles 
of the \ion{Fe}{i} 5379.5740~{\AA} line for three different line 
strengths computed by changing the oscillator strengths, $\log gf$, in the 
3D spectral synthesis by -0.1, 0.0 and +0.1~dex, respectively. 
We ran 3D spectral synthesis for a subsample of 205 Fe lines, with 112 and 93 lines in 
the blue and in the red chips, respectively. 
The wavelength centroid of the core of these lines were measured using a 
similar approach as for the observed line cores. 
Figure~\ref{fig:lc3D} shows the position of the
line cores of three different 3D profiles for the three $gf$ values with 
EWs of 53.2, 62.1 and 71.0~\ma, providing a theoretical core shift of 
$-345.3$, $-280.9$ and $-215.2$~\ms, respectively, for these three 3D profiles
with respect to the laboratory wavelength of the \ion{Fe}{i} 5379.5740~{\AA} line.
The theoretical core shift (always negative since there is no solar GRS added
in the 3D model) gets closer to zero as the line gets stronger, so
the convective blue shifts is smaller as the lines gets stronger.
We performed a linear interpolation with the observed EW to infer a line core shift 
of $-294.6$~\ms~for the \ion{Fe}{i} 5379.5740~{\AA} line, with an EW 
of $60.2\pm1.2$~m{\AA}.
The 3D model provides a qualitative and quantitative description of the granulation 
phenomena, where the weaker the lines are the more blue shifted, as expected. 
Subtracting the line core shift derived from the 3D line profile, 
$v_{\rm core,3D}=-294.6$~\ms~to the observed line core shift of the observed 
line \ion{Fe}{i} $\lambda_{\rm lab}$ 5379.5740~{\AA}, 
$v_{\rm core,obs} =346.6\pm18.4$~\ms, should provide a value of the solar 
gravitational redshift, $v_{\rm GRS,obs} = 641.2\pm18.4$~\ms, 
which is in agreement with the theoretical value of 633.10~\ms. 

\section{3D global line shifts \label{sec:3dlfit}}

We also performed the global fit of the observed line profiles using the 3D synthetic 
profiles. For each observed line, a grid of 3D synthetic profiles is created for 
three different values of Fe abundance (A(Fe)$= 7.42$, 7.52 and 7.62)
and rotational broadening ($v_{\rm ROT} = 0$, 2, and 4~\kms).
The grid is first broadened using the code {\sc LinforRotate} \citep{lud07} 
and later with a Gaussian instrumental profile with a full width half maximum (FWHM)
equivalent to the resolving power $R=115,000$ ($v_{\rm INS}=2.61$~\kms). 
We performed a global fit of every Fe line using our automated IDL-based routine that 
makes use of the MPFIT\footnote{http://purl.com/net/mpfit} routine~\citep{mar09}, 
including three free parameters: rotational broadening, velocity shift, and iron 
abundance. The continuum location was fixed at 1.
We ran this routine to fit the subsample of Fe lines for the five HARPS-LFC spectra.

In Fig.~\ref{fig:l3Dfit}, we show  
an example of the global line fit to the 
\ion{Fe}{i} $\lambda_{\rm lab}$ 5379.5740~{\AA} line.
The fit is done in a wavelength range of 0.5~{\AA}
centred on the laboratory wavelength. The 3D synthetic profile matches quite well 
the observed line profile, although, as shown in the bottom-right panel of
Fig.~\ref{fig:l3Dfit}, the 3D synthetic profile does not perfectly fit the 
line core. From the fit of this spectral line in the spectrum s429 
(see Table~\ref{tab:obs}), we infer a global velocity shift of 
$v_{\rm fit,3D} = 648.4\pm4.8$~\ms~with a solar abundance of 
A(Fe)$=7.526\pm0.001$~dex and $v_{\rm ROT} = 1.72\pm0.06$~\kms. Both the iron 
abundance and the rotational velocity are consistent with their expected solar values.  
The expected rotational velocity
of the Sun is $v\sin i=1.8\pm0.2$~\kms~\citep[see e.g.][]{gra18}. 
The mean rotation and Fe abundance values found for the final set of fitted Fe lines 
are 2.5~\kms\ and 7.59~dex with a standard deviation of 0.5~\kms\ and 0.09~dex.
The derived velocity shift is close although not consistent with regard to the theoretical 
value of the solar gravitational redshift, $v_{\rm GRS,theo}=633.1$~\ms.
The result depends on the accuracy of the laboratory wavelength, with an uncertainty
of $> 16$~\ms~for lines labelled as A, as well as the precision of the 3D profile,
with an uncertainty of about 30~\ms. Hence, a deviation $\sim 15$~\ms~is within 
this error.

\begin{figure}
\begin{center}
{\includegraphics[clip=true,width=95mm,angle=0]{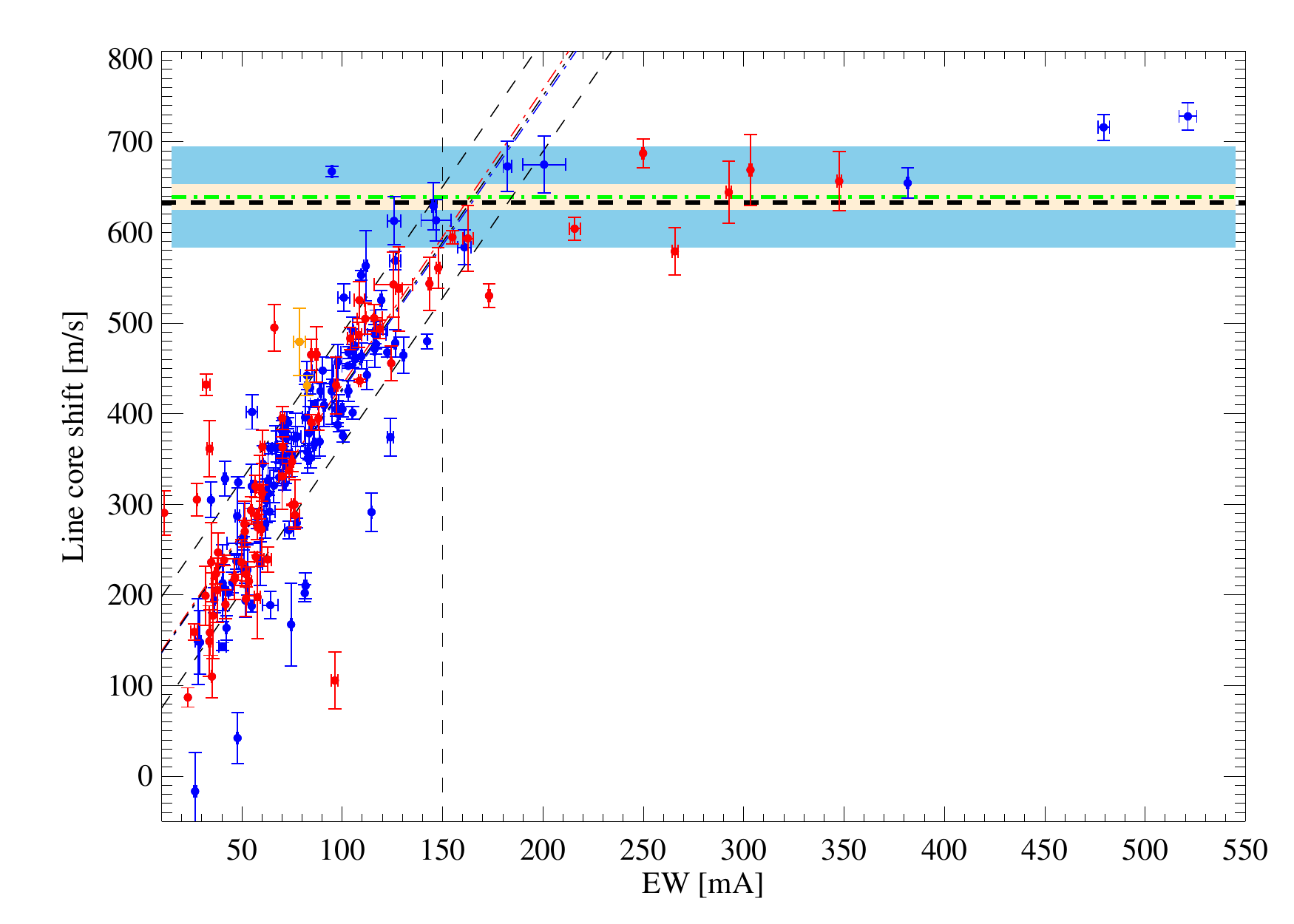}}
\end{center}
\caption{
Line core shift of \ion{Fe}{i} spectral lines, 
estimated using the recalibrated 
wavelengths $\lambda_{\rm nist}$ as reference laboratory wavelengths,
measured on HARPS-LFC 
Moon spectra in the blue chip (blue dots) and in the red chip (red dots). 
Line core shifts of the two \ion{Fe}{ii} lines are displayed as orange symbols.
Blue, red, and black dashed-dotted lines show the linear fits of the blue, 
red, and all dots corresponding to the lines with EWs smaller then 170~m{\AA}. 
The horizontal black dashed line shows the theoretical solar gravitational 
redshift (GRS) measured on Earth, $v_{\rm GRS,theo}=633.10$~m~s$^{-1}$.
The green dashed-dotted line shows the mean line core shift, $v_{\rm GRS,obs}$, 
of the $N=15$ lines at EWs greater than 150~m{\AA}.
The light blue and light yellow regions show the standard deviation, 
$\sigma_{\rm GRS,obs}$, and the $\sigma_{\rm GRS,obs}/\sqrt{N}$.
\label{fig:grs_lc}
}
\end{figure}

\section{Solar gravitational redshift \label{sec:grs}}

We computed the mean line core shifts of each line of the sample of 257 Fe lines 
measured on the five HARPS-LFC spectra of the Moon. The uncertainties are computed as
the standard deviation of the measurements. In order to remove outliers in the 
measurements among the line core shifts of the five spectra, 
we rejected all the lines with an line core shifts different from the mean by more 
than a factor of 1.5 times the standard deviation. 
We checked other threshold factors such as 2, 2.5, and 3 times the standard deviation, 
but we decided to remain in a restrictive position. 
This factor of 1.5 versus 2.0 or 3.0 does not have a significant 
implication (i.e. $\leq 5-10$~\ms~difference) for the mean line core shift, but in some 
cases, it has an appreciable effect on the final uncertainty. 

In Fig.~\ref{fig:grs_lc}, we display the mean line core shifts with respect 
to their equivalent widths. 
We identify 12 and 15 lines in the blue and red chips with uncertainties on line core 
shift between 50 and 100~\ms~and with EWs smaller than 60~\ma, which were discarded from 
Fig.~\ref{fig:grs_lc}. 
%We then select those lines with laboratory wavelength flagged as A.
Among a total of 230 lines, that is, 141 and 89 lines in the blue and red chips, respectively, 
a final sample of 188 lines are depicted in Fig.~\ref{fig:grs_lc}, 
with 116 and 72 in the blue and red, which have laboratory wavelengths flagged as A. 
%In Table~\ref{tab:lc}, we provide the original wavelengths, $\lambda_{\rm lab}$, 
In Table~A.1, we provide the original wavelengths, $\lambda_{\rm lab}$, 
and the recalibrated wavelengths, $\lambda_{\rm nist}$ and $\lambda_{\rm ritz}$, 
together with other line parameters and with the observed line core shifts, 
$v_{\rm core,obs}$, of these 188 lines, computed as explained in Section~\ref{sec:lc}. 
This line core shift uses as reference wavelength $\lambda_{\rm lab}$. 
Using the $\lambda_{\rm nist}$ and $\lambda_{\rm ritz}$ would lead to a greater 
line core shift by about 20 and 35~\ms, respectively.

The measured line core shift, $v_{\rm core,obs}$, of these lines increases towards 
larger EWs up to about 170~\ma\ and then stays roughly constant.
As this increase seems to be linear, we fit the line core shifts versus EWs with 
linear functions for both the blue and the red lines and with all lines together 
and we find that this linear fits intersects the theoretical solar gravitational 
redshift at about 170~\ma. 
We fit line core shifts versus equivalent widths for lines weaker than 150~\ma. 
The standard deviation of the residuals of this fit is roughly 61~\ms, whereas 
using the $\lambda_{\rm nist}$ and $\lambda_{\rm ritz}$ gives a standard deviation 
of the fit of 61~\ms\ and 67~\ms, respectively. 
In Fig.~\ref{fig:grs_lc}, we depict the result using $\lambda_{\rm nist}$ as the
reference laboratory wavelengths. The fit of lines weaker than 150~\ma\ intersects
the theoretical GRS at $\sim 165$~\ma and using the standard deviation of the fit 
gives an uncertainty of $\sim 15$~\ma. The observed line core shifts for lines
stronger than approximately 150~\ma\ stays roughly constant, reaching a similar value
as the theoretical solar GRS.

Theory predicts that convective effects on the photospheric lines are less 
important as their cores are formed higher in the solar atmosphere. 
Lines with larger EWs, and therefore stronger lines, have their cores formed 
higher in the atmosphere. 
These observations seem to indicate that above a certain EW limit 
(see Fig.~\ref{fig:grs_lc}), the 
lines do not show any convective blue shift and, thus, their cores are only 
affected by the solar gravitational redshift.
Therefore, we consider strong lines with EWs larger than 150~\ma~appropriate 
for observational measurements of the solar gravitational redshift. 
There are 15 strong lines
with $150 <$~EW[\ma]~$< 550$, that is, 6 and 9 lines in blue and red chips, respectively.
We computed the mean line core shift for these strong lines at 
$v_{\rm GRS,obs} = 639\pm14$~\ms.
The uncertainty is derived as the standard 
deviation of the measurements, $\sigma_{\rm GRS,obs}$, from the mean 
divided by the square root of the number of measurements, $N$, and thus, the 
final uncertainty as $\delta v_{\rm GRS,obs} = \sigma_{\rm GRS,obs} / \sqrt{N}$.
This measurement is consistent with the theoretical 
value of the solar gravitational redshift, $v_{\rm GRS,theo}=633.1$~\ms. As an 
observational test of the general theory of relativity~\citep{ein16}, 
it represents the most robust spectroscopic measurement of the gravitational 
redshift of the Sun.

\begin{figure}
\begin{center}
{\includegraphics[clip=true,width=95mm,angle=0]{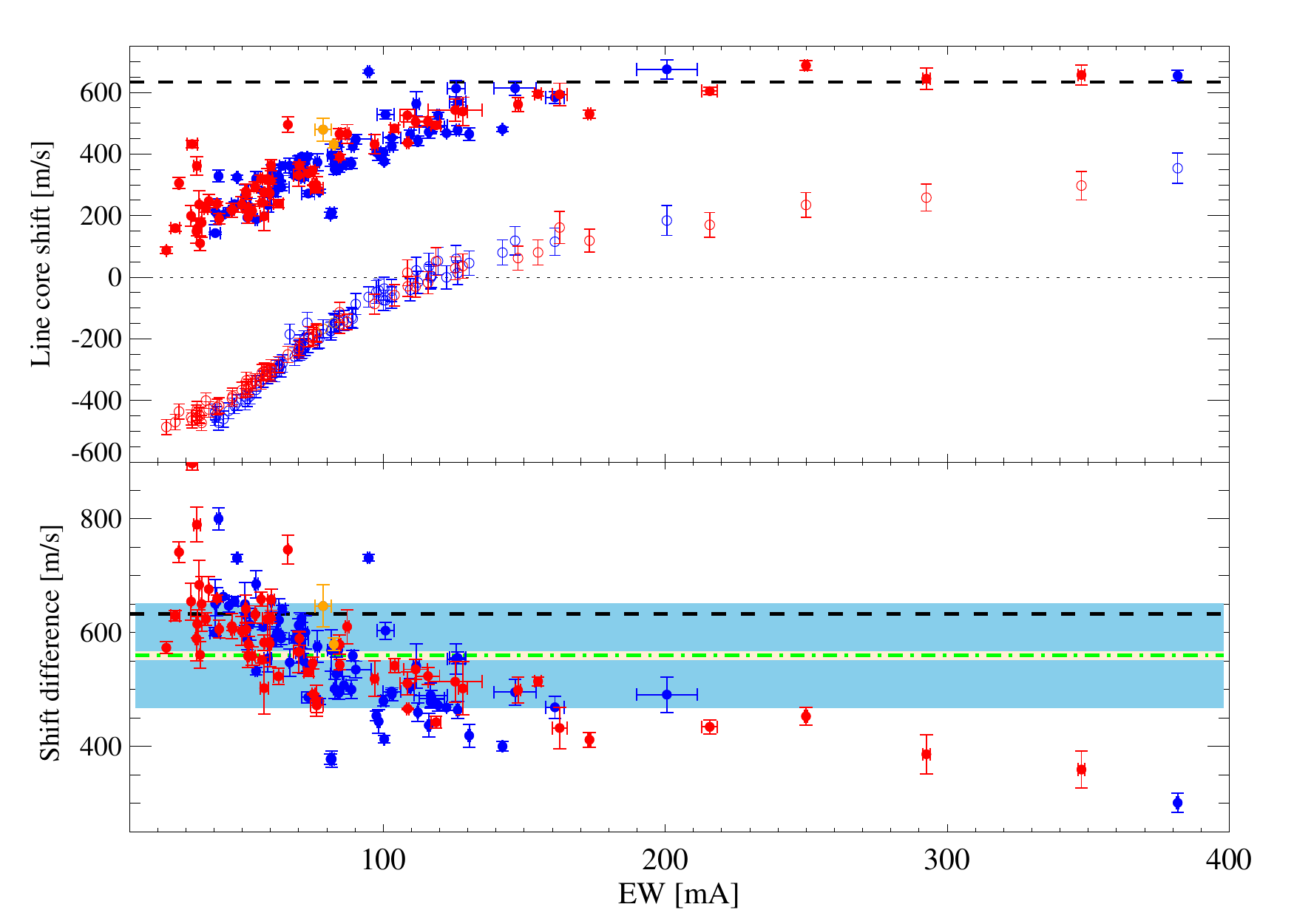}}
\end{center}
\caption{
{\it Top panel}: Line core shift of \ion{Fe}{i} spectral lines, 
estimated using the recalibrated 
wavelengths, $\lambda_{\rm nist}$, as reference laboratory wavelengths,
measured on HARPS-LFC Moon spectra in 
the blue chip (blue dots) and in the red chip (red dots), and 
those measured on 3D hydrodynamical theoretical line 
profiles (open circles) for the same lines. 
Line core shifts of the two \ion{Fe}{ii} lines are displayed as orange symbols.
The dotted line shows the zero shift for reference.
{\it Bottom panel}: Line core shift difference of the observed spectral 
lines and the theoretical line profiles.
The black dashed line in both panels shows the theoretical solar GRS.
The green dashed-dotted line shows the mean line core shift, 
$v_{\rm GRS,3D,lc}$, of the $N=144$ lines at $10 < EW[\ma] < 400$.
The light blue and light yellow regions show the standard deviation, 
$\sigma_{\rm GRS,3D,lc}$, and the $\sigma_{\rm GRS,3D,lc}/\sqrt{N}$.
\label{fig:grs_lc3D}
}
\end{figure}

In Fig.~\ref{fig:grs_lc3D} we display both the observed and theoretical line core 
shifts of lines with $10 <$~EW[\ma]~$< 400$. A total of 144 lines 
is depicted, that is, 80 and 64 in the blue and red chips, respectively, with laboratory wavelengths 
flagged as A. Again, we chose the recalibrated wavelengths, $\lambda_{\rm nist},$ 
as reference wavelengths.
The line core shift of the 3D profiles are estimated by linearly interpolating 
the line core shifts of the three line profiles in the observed EW according to 
the EW of the 3D profiles. 
The uncertainties on the line shifts of 3D profiles from different 
snapshots are estimated as the standard deviation of the line 
core shifts of the different snapshots divided by the 
square root of the number of snapshots, $N=20$.
The uncertainty versus EW (and therefore line depth) increases 
progressively from 25~\ms\ at EW~$=20$~\ma\ to about 50~\ms\ at EW~$=400$~\ma. 
The mean uncertainty is 31~\ms.

The difference between the predicted line core (red) shift of the lines with 
the observed spectra of the Moon and the line core (mostly blue) shift 
measured in the 3D line profiles should also provide another observational 
measurement of the solar gravitational redshift. 
In the top panel of Fig.~\ref{fig:grs_lc3D}, we can see that both the observed and
theoretical line core shifts increase towards larger EWs. Qualitatively the 3D 
line cores follow the same behaviour as the observed line cores. However,
as seen in the bottom panel of Fig.~\ref{fig:grs_lc3D},  
the velocity field of the 3D model atmosphere does not exactly reproduce 
the observations quantitatively, as the 3D line cores are more redshifted than the observed 
line cores as the lines are stronger than about 60~\ma. 
Stronger lines are more redshifted than expected, and at about 110~\ma,~the line core 
shift becomes positive and continues to increase towards larger equivalent widths.
This effect has been previously reported in the literature for both 
CO$^5$BOLD and STAGGER models~\citep[e.g.][]{all09}.

The cores of strong lines form at heights that can be influenced by the 
upper boundary of the 3D model. 
%The solar model for which we performed the 
%spectral synthesis calculations exhibits a systematic outflow of mass 
%corresponding to a velocity a few 10~\ms~in vicinity of the upper boundary. 
%This means that the redshifted cores of the strong lines cannot be related to 
%this numerical artifact since an outflow would rather produce blueshifted cores. 
There must be some effects missing in the 3D model and, thus, subsequent 
spectral synthesis calculations, which lead to a decorrelation between velocity 
and temperature in the upper regions of the photosphere that would remove 
line shifts. Candidates are non-LTE effects and magnetic fields.  
While conducting non-LTE spectral syntheses for the many iron lines 
is beyond the scope of this paper, test calculations including magnetic 
fields in 3D models show that the presence of magnetic fields indeed reduces 
the redshifts of line cores. We could not obtain a quantitative correspondence. 
Moreover, magnetic fields that are strong enough to reduce the core shift of 
strong lines also reduce the shift of weak lines. 
Thus, magnetic field tend to suppress both negative and positive intrinsic 
line shifts both in weak and strong lines, respectively. 
A similar result has been found in \citet{per13}, as seen in their Fig.~14 
for lines weaker than 100~\ma.
While the field geometry in local-box 3D models is somewhat arbitrary, 
it is difficult to envision a configuration that reduces the shift of strong 
lines while retaining the shift of weak lines. 
Non-LTE effects may be the reason for the mismatch.
In 3D NLTE, the over-ionization of \ion{Fe}{i} reduces line strengths changing 
the formation depth, in particular, low-excitation are significantly 
weakened, possibly affecting line shifts~\citep{hol13,lin17}. 

The difference between observed line core shifts and 3D line core shifts  
shows that weak lines with EWs smaller than 60~\ma~appear to show consistency 
with the solar GRS. 
However, the difference gets larger as we increase more and more the strengths 
of the lines. 
The mean shift difference between observed and 3D line cores
is $v_{\rm GRS,3D,lc}=560\pm8$~\ms. The uncertainty is again derived as 
$\sigma_{\rm GRS,3D,lc}/\sqrt{N}$ with the standard deviation,  
$\sigma_{\rm GRS,3D,lc}$, and the number of lines, $N=144$ 
(see Fig.~\ref{fig:grs_lc3D}).

The line cores of 3D profiles do not exactly follow the same behaviour as the 
observed line cores, but 3D line profiles, as a whole, may do be shown to do so. 
We explored this possibility by computing global line shifts to the observed
lines with the 3D profiles using our own automated code, as shown 
in Section~\ref{sec:3dlfit}.

\begin{figure}
\begin{center}
{\includegraphics[clip=true,width=95mm,angle=0]{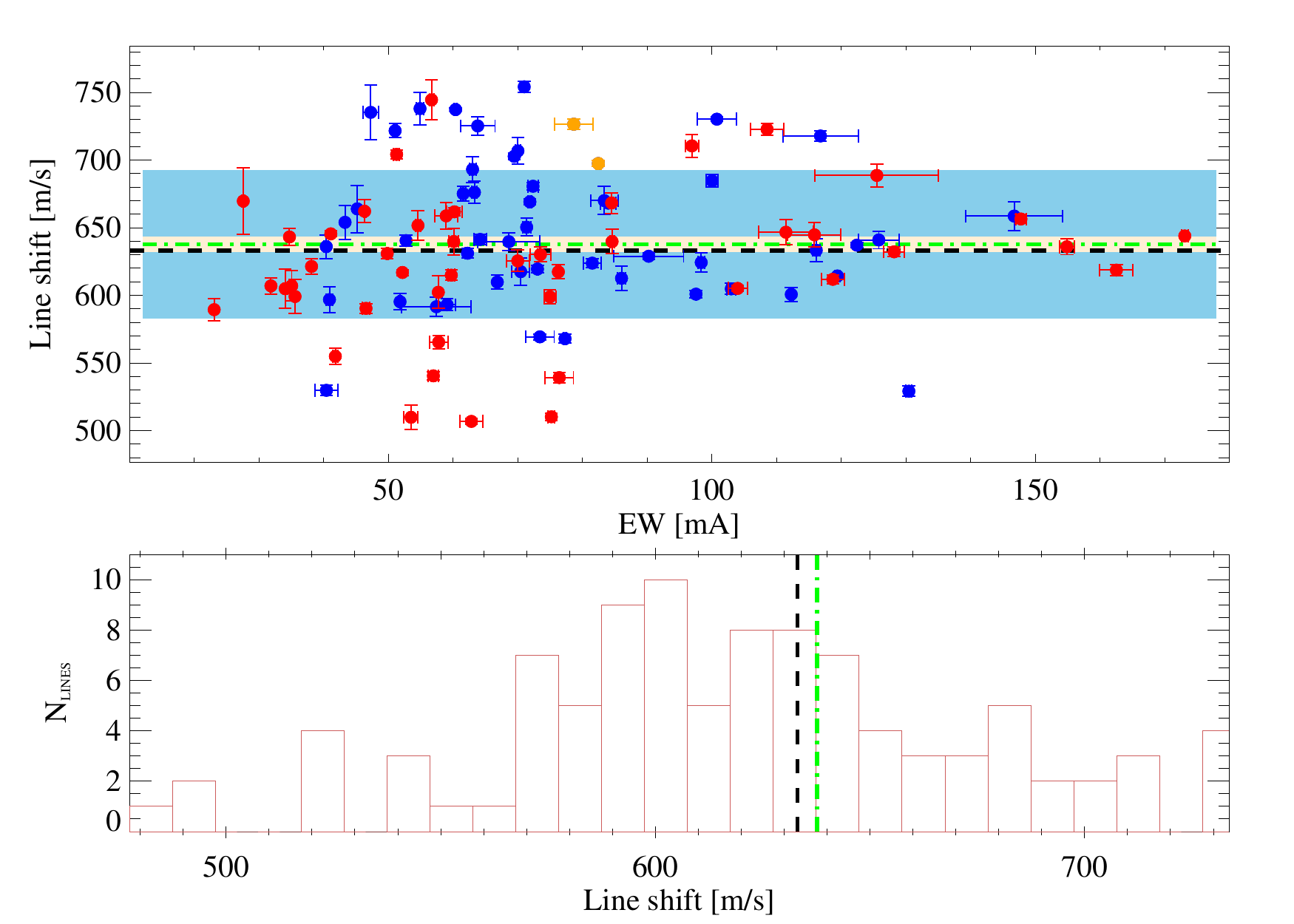}}
\end{center}
\caption{
{\it Top panel}: Line shift from \ion{Fe}{i} spectral lines, 
estimated using the recalibrated 
wavelengths, $\lambda_{\rm nist}$, as reference laboratory wavelengths,
fitted on HARPS-LFC Moon spectra with 3D line profiles in the blue chip 
(blue dots) and in the red chip (red dots).
Line shifts of the two \ion{Fe}{ii} lines are displayed as orange symbols.
The light blue and light yellow regions show the standard deviation, 
$\sigma_{\rm GRS,3D}$, and the $\sigma_{\rm GRS,3D}/\sqrt{N}$.
{\it Bottom panel}: Histogram of the line shifts.
The black dashed line shows the theoretical solar GRS.
The green dashed-dotted line shows the mean line shift, $v_{\rm GRS,3D}$, 
of $N=97$ lines at $10 < EW[\ma] < 180$.
\label{fig:grs_lf3D}
}
\end{figure}

In Fig.~\ref{fig:grs_lf3D}, we display global line shifts, estimated using the recalibrated 
wavelengths $\lambda_{\rm nist}$ for the reference laboratory wavelengths, as a function of the observed EW of the Fe lines. 
We discarded lines by visually inspecting the global line fits and by computing 
the reduced $\chi$ square, 
$\chi^2_\nu=\sum_{i=1}^{n}[(f_{i{\rm ,obs}}-f_{i{\rm ,3D}})/\sigma_i]^2 / \nu $, 
where $i$ corresponds to each of the $n=11$ observed flux points, $f_{i{\rm ,obs}}$,
around the line core, with their uncertainties, $\sigma_i$, and the 3D profile flux 
values, $f_{i{\rm ,3D}}$, and $\nu=n-m$ are the degrees of freedom and $m=3$ is the
number of fitted parameters (see bottom-right panel of Fig.~\ref{fig:l3Dfit}).
As the good line fits, we adopt those with apparent good global line fits and reasonable 
line core fits, with $\chi_\nu^2 < 25$.  
We apply a 3-$\sigma$ clipping procedure on the global line shifts to remove five 
lines with shifts above 800~\ms, thus leaving a final number of 97 lines
with $10 <$~EWs[\ma]~$< 180$, 51 and 46 lines
in the blue and red chips, respectively.
We decided to restrict the upper boundary on EW of the global line fits 
to 180~\ma~to be able to compare the results with those of line core shifts. 
In addition, stronger lines are more difficult to fit globally since these lines 
are typically blended with other lines and a proper fit should include these 
weaker lines to be able to faithfully reproduce the line profiles. 
 
The observed line core shifts, $v_{\rm core,obs,n}$, computed using the recalibrated 
wavelengths, $\lambda_{\rm nist}$, as reference laboratory wavelengths, 
the 3D line core shifts, $v_{\rm core,3D}$, and the global line shifts, 
$v_{\rm fit,3D,n}$, corrected using the recalibrated wavelengths $\lambda_{\rm nist}$ as 
%reference laboratory wavelengths, of these 97 lines are provided in Table~\ref{tab:lc3d} 
reference laboratory wavelengths, of these 97 lines are provided in Table~A.2
together with other line parameters.
In Fig.~\ref{fig:grs_lf3D}, these global line shifts, $v_{\rm fit,3D,n}$, are 
distributed around the expected solar GRS.
The mean shift of observed lines with respect to the 3D line profiles 
is $v_{\rm GRS,3D}=638\pm6$~\ms, in agreement with the 
theoretical value of the solar GRS.
The uncertainty is again determined as $\sigma_{\rm GRS,3D}/\sqrt{N}$ with $N=97$ 
lines.

\section{Discussion and Conclusions}

The analysis of high-quality HARPS spectra of the Moon calibrated with the laser 
frequency comb allows us to obtain an observational measurement of the solar GRS.
We performed the analysis on an automated basis, using data reduction, 
wavelength calibration, continuum normalisation, and identification of lines and 
measurement of equivalent widths, line core shifts and global line shifts. 
There are uncertainties hidden in these process that certainly introduce some 
scatter in the measurements at the level of several tens of \ms. 
We note the pixel size is $\sim 800$~\ms~in the HARPS blue and red detectors. 
We are able to show a rather clear boundary in the equivalent widths of the 
lines,  
at $165\pm15$~\ma, where lines with larger equivalent widths do not show 
any sign of convective shift. 
The 15 Fe lines with EWs larger than 150~\ma, are distributed 
around the theoretical value of solar GRS with a dispersion of 56~\ms. 
The mean uncertainty on line core shift of these 15 lines is 23~\ms,
 which is almost half of the dispersion of the measurements. 
This may be explained by the uncertainty on laboratory wavelengths at the 
level of $16-75$~\ms~\citep{nav94}. 
The mean uncertainty on EW for these 15 Fe lines is 2.4~\ma.
For the total number of 188 Fe lines, the mean uncertainty on line core shift is 
19.2~\ms, whereas mean uncertainty on EW is only 1.2~\ma.

The velocity field in the 3D model atmosphere of the Sun seems to reproduce 
quite well the observed behaviour of line core shifts, at least qualitatively, 
and quantitatively for weak lines with EW~$< 60$~\ma. 
On the other hand, the global line shift from the 3D profiles matching the 
observed profiles behaves better although there remains a dispersion in the 
measurements around the mean of about 73~\ms~for the sub-sample of 102 Fe lines, 
and goes down to 55~\ms~if we discard the five outliers by applying a 
$\sigma$ clipping procedure, leaving 97 Fe lines depicted in Fig.~\ref{fig:grs_lf3D}.
The mean uncertainty on the global line shift that comes from the automated fitting 
procedure and computed as the standard deviation from the measurements of the 
five HARPS-LFC spectra is much smaller at 6~\ms. However, the statistical 
uncertainty on the wavelength position of the average 3D profiles is a bit larger
at about 30~\ms.
This may indicate that the dispersion around the theoretical value of solar 
GRS of 55~\ms~may be indeed related to the uncertainty on the laboratory 
wavelengths.
The recalibrated wavelengths, $\lambda_{\rm nist}$, do provide a slightly 
different result than the original laboratory wavelengths, $\lambda_{\rm lab}$, 
in \citet{nav94}, shifting the result by +20~\ms.
Similarly, although the uncertainties on the recalibrated Ritz 
wavelengths estimated from the energy levels, $\lambda_{\rm ritz}$, are smaller 
than those of the original wavelengths, using the $\lambda_{\rm ritz}$ moves 
the result by +35~\ms\ and provides a larger dispersion.
New laboratory experiments to improve the accuracy and revise the 
laboratory wavelengths of \ion{Fe}{i} and other element species abundant in the 
visible solar spectrum appears necessary.

Assuming the recalibrated wavelengths, $\lambda_{\rm nist}$, as a laboratory reference, 
we have been able to achieve an observational measurements of the solar GRS of 
$v_{\rm GRS,3D}=638\pm6$~\ms~from the mean of observed global line shifts of 
97 Fe lines with $10 <$~EWs[\ma]~$< 180$, and 
$v_{\rm GRS,obs} = 639\pm14$~\ms~from the mean 
line core shift of 15 strong Fe lines with EW~$> 150$~\ma.
Both measurements are in perfect agreement with the theoretical value of the 
solar gravitational redshift, $v_{\rm GRS,theo}=633.1$~\ms, representing 
an observational test of the general theory of relativity~\citep{ein11,ein16}.
At the same time, our observations point out the high quality, along with some 
limitations, in the current 3D modelling of the solar lines.
New, high-quality spectra of the Moon taken with HARPS~\citep{may03}, or at
an even higher resolution with ESPRESSO~\citep{pep14,gon18} and calibrated with the 
laser frequency comb in a wider spectral range, could provide a larger number of Fe 
lines that can be used to measure accurate line shifts (and, in fact, line bisectors) 
as diagnostics for understanding the structure and dynamics of the solar photosphere and for 
validating and improving 3D model atmospheres,
in addition to serving as a tool for definitively probing solar gravitational redshift. 

\begin{acknowledgements}
JIGH acknowledges financial support from the Spanish Ministry of Science and Innovation 
(MICINN) under the 2013 Ram\'on y Cajal program RYC-2013-14875. JIGH, RRL, ASM and BTP 
acknowledge financial support from the Spanish ministry project MICINN AYA2017-86389-P.
ASM acknowledges financial support from the Spanish MICINN under the 2019 Juan de la 
Cierva Programme. BTP acknowledges Fundaci\'on La Caixa for the financial support 
received in the form of a Ph.D. contract. 
EC gratefully acknowledge support from the French National Research Agency (ANR) 
funded project ``Pristine'' (ANR-18-CE31-0017).
HGL acknowledges financial support by the Deutsche Forschungsgemeinschaft
(DFG, German Research Foundation) -- Project-ID 138713538 -- SFB 881
(``The Milky Way System'', subproject A04).
This work has made use of the VALD database, operated at Uppsala University, 
the Institute of Astronomy RAS in Moscow, and the University of Vienna.
This work also used IRAF package, which are distributed by the National Optical 
Astronomy Observatory, which is operated by the Association of Universities for 
Research in Astronomy, Inc., under contract with the National Science Foundation.
This work makes also use of the NIST Atomic Spectra Database at
National Institute of Standards and Technology, Gaithersburg, MD.
\end{acknowledgements}


\bibliographystyle{aa} % style aa.bst
\bibliography{lfcharps} % your references Yourfile.bib

\end{document}